\documentclass[11pt,twoside]{article}
\usepackage{asp2010}
\def\Msol{\mbox{ }M_{\odot}}
\def\Lsol{\mbox{ }L_{\odot}}
\def\Rsol{\mbox{ }R_{\odot}}
\def\Tstar{\mbox{ }T_{\star}}
\def\Rstar{\mbox{ }R_{\star}}
\def\ergs{\mbox{ erg\,s}^{-1}}
\def\kms{\mbox{ km\,s}^{-1}}
\def\amin{^\prime}
\def\adeg{^{\circ}}
\def\nh{N_{\rm H}}
\def\cmmoinsdeux{\mbox{ cm}^{-2}}
\def\mags{\mbox{ magnitudes}}
\def\microns{\mbox{ } \mu \mbox{m}}
\def\igrjstu{\mbox{IGR~J16318$-$4848}}

\resetcounters

\begin{document}

\title{Nature, formation and evolution of High Mass X-ray Binaries}
\author{Sylvain~Chaty
\affil{Laboratoire AIM (UMR 7158 CEA/DSM-CNRS-Universit\'e Paris Diderot),
Irfu/Service d'Astrophysique, CEA-Saclay, FR-91191 Gif-sur-Yvette Cedex, France, chaty@cea.fr}}

\begin{abstract}
The aim of this review is to describe the nature, formation and evolution of the three kinds of high mass X-ray binary (HMXB) population: {\it i.} systems hosting Be stars (BeHMXBs), {\it ii.} systems accreting the stellar wind of supergiant stars (sgHMXBs), and {\it iii.} supergiant stars overflowing their Roche lobe.
There are now many new observations, from the high-energy side (mainly from the {\it INTEGRAL} satellite), complemented by multi-wavelength observations (mainly in the optical, near and mid-infrared from ESO facilities), showing that a new population of supergiant HMXBs has been recently revealed. New observations also suggest the existence of evolutionary links between Be and stellar wind accreting supergiant X-ray binaries.
I describe here the observational facts about the different categories of HMXBs, discuss the different models of accretion in these sources (e.g. transitory accretion disc versus clumpy winds), show the evidences of a link between different kinds of HMXBs, and finally compare observations with population synthesis models.
\end{abstract}

\section{Introduction}

Nearly 50 years after the discovery of the first extra-solar X-ray source \citep[Sco\,X-1;][]{giacconi:1962}, X-ray astronomy has now reached its age of reason, with a plethora of telescopes and satellites covering the whole electromagnetic spectrum, obtaining precious observations of these powerful celestial objects.


High energy binary systems are composed of a compact object -- neutron star (NS) or black hole (BH) -- orbiting a companion star, and accreting matter from it \citep[cf e.g.][for a review]{chaty:2006a}. The companion star is either a low mass star (typically $\sim 1 \Msol$ or less, with a spectral type later than B, called in the following LMXB for ``Low-Mass X-ray Binary''), or a luminous early spectral type OB high mass companion star (typically $> 10 \Msol$, called HMXB for ``High-Mass X-ray Binary''). 300 high energy binary systems are known in our Galaxy: 187 LMXBs and 114 HMXBs \citep[respectively 62\% and 38\% of the total;][]{liu:2006,liu:2007}.

Accretion of matter is different for both types of sources\footnote{I will not review here the so-called $\gamma$-ray binaries, where most of the energy is in the GeV-TeV range, coming from interaction between relativistic electrons of the pulsar and the wind of the massive star.}. In the case of LMXBs, the small and low mass companion star fills and overflows its Roche lobe, therefore accretion of matter always occurs through the formation of an accretion disc. The compact object can be either a NS or a BH, Sco X-1 falling in the former category.
In the case of HMXBs, accretion can also occur through an accretion disc, for systems in which the companion star overflows its Roche lobe; however this is generally not the case, and there are two alternatives. The first one concerns stars with a circumstellar disc, and here it is when the compact object -- on a wide and eccentric orbit -- crosses this disc, that accretion periodically occurs (case of HMXBs containing a main sequence early spectral type Be III/IV/V star, rapidly rotating, called in the following BeHMXBs). The second case is when the massive star ejects a slow and dense wind radially outflowing from the equator, and the compact object directly accretes the stellar wind through e.g. Bondy-Hoyle-Littleton processes (case of HMXBs containing a supergiant I/II star, called sgHMXBs).
We point out that there also exists binary systems for which the companion star possesses an intermediate mass (typically between 1 and $10 \Msol$, called IMXB for "Intermediate-Mass X-ray Binary").

\section{Different types of High Mass X-ray Binaries}

\subsection{Be X-ray binaries}

BeHMXBs usually host a NS on a wide and eccentric orbit around an early spectral type B0-B2e (with emission lines) donor star, surrounded by a circumstellar ``decretion'' disc of gas created by a low-velocity and high-density stellar wind of $\sim 10^{-7} \Msol / yr$. This disc is characterized by an H$\alpha$ emission line (whose width is correlated with the disc size) and a continuum free-free or free-bound emission \citep[causing an infrared excess;][]{coe:2000, negueruela:2004}\footnote{For more details about BeHMXBs, I warmly recommend the excellent reference chapter by \cite{charles:2006} on optical/infrared emission of HMXBs.}.
They exhibit transient and bright X-ray outbursts: {\it i.} ``Type I'' are regular and periodic each time the NS crosses the decretion disc at periastron; {\it ii.} ``Type II'' are giant outbursts at any phase, with a dramatic expansion of the disc, enshrouding the NS; {\it iii.} ``Missed'' outbursts exhibit low H$\alpha$ emission (due to a small disc or a centrifugal inhibition of accretion), iv. ``Shifting outburst phases'' are likely due to the rotation of density structures in the circumstellar disc.

There are $\sim 50$ BeHMXBs in our Galaxy, and $>35$ in the Small Magellanic Cloud (SMC). This large number in the SMC, instead of $\sim 3$ predicted by the galactic mass ratio $\frac{Milky~Way}{SMC} \sim 50$, is likely due to the bridge of material between our Galaxy and the Magellanic Clouds \citep{mcbride:2008}. There is a large number of Supernova Remnants (SNRs) of similar age ($\sim 5$\,Myr) suggesting an increased starbirth rate due to tidal interactions \citep{stanimirovic:1999}, since the previous closest SMC/Large Magellanic Cloud (LMC) approach was $\sim 100$\,Myr ago, and new massive stars probably formed the current HMXB population. 
A strong spatial correlation exists between emission line stars and stars aged $8-12$\,Myr, with BeHMXBs in the SMC \citep{meyssonnier:1993, maragoudaki:2001}. 
The number of HMXBs is therefore an indicator of star formation rate and starburst activity \citep{popov:1998, grimm:2003}, and the Magellanic Clouds provide an uniform sample of BeHMXBs in a compact region, all at the same distance.

\subsubsection{Formation of Be X-ray binaries}

BeHMXB populations are similar in the Milky Way and SMC/LMC, when examining the spectral type of the companion star \citep{negueruela:2002, mcbride:2008}. However, while comparing the spectral type of Be stars hosted in X-ray binaries versus isolated Be, all located in our Galaxy, it appears that the spectral type distribution of Be in X-ray binaries is much narrower, beginning at O9 V ($\sim 22 \Msol$), peaking at B0 V ($\sim 16 \Msol$) and stopping at B2 V ($\sim 8-10 \Msol$). The mass distribution of Be stars, ranging between 8 and $22 \Msol$, is consistent with the fact that wide orbits are vulnerable to disruption during SN event, especially for less massive B stars. The low mass range therefore supports the existence of SN kick velocities and angular momentum loss, low mass stars being ejected, while heavier systems become sgHMXBs \citep{portegies-zwart:1995, mcbride:2008}.

The formation of BeHMXBs has been explained by the so-called model of ``rejuvenation'' \citep{rappaport:1982}, by being a product of binary evolution. In this model, the mass transfer speeds up first the rotation of the outer layers of the B spectral type secondary star, and then the star itself, which starts to turn so quickly that a circumstellar disc of gas is created, giving birth to the Be phenomenon \citep[i.e. these stars are neither born as fast rotators, nor spun-up in the final main-sequence stages;][]{coe:2000, charles:2006}.
These systems are thus the result of moderately massive binaries undergoing a semi-conservative mass transfer evolution, with wide orbits (200 to 600\,days) produced before the first SN event of the system, and eccentric orbits due to small asymmetries during the SN event \citep{vandenheuvel:1983, verbunt:1995}. This model is consistent with the spectral types of the companion stars between O9 and B2V: stars less massive than $\sim 8 \Msol$ are ejected during the SN event, and stars more massive than $\sim 22 \Msol$ become supergiant stars.

\subsubsection{Structure of Be circumstellar disc}

In BeHMXBs, Be circumstellar discs are naturally truncated due to the presence of the NS, as a result of tidal torques at certain resonance points (where the Keplerian period $P_K$ is an integer fraction of the orbital period $P_\mathrm{orb}$), thus preventing any accretion of matter beyond these points. 
The accretion on the NS is unlikely for BeHMXBs where the NS is on a circular orbit, with a disc always truncated at a fixed size, smaller than the Roche lobe. This creates a persistent low-level X-ray emission from the stellar wind, and occasional Type II outbursts.
Instead, BeHMXBs with a NS on a highly eccentric orbit will allow for periodic accretion, with the size of the rotating truncated disc depending on the orbital phase, and at periastron the disc can include the NS orbit, thus creating Type I X-ray outbursts \citep{okazaki:2001, reig:2007}.

In addition, recent observations show that circumstellar discs in BeHMXBs also exhibit cycles of activity, likely due to their formation and dispersion. BeHMXBs therefore exhibit 3 periods: a spin period $P_\mathrm{spin}$, an orbital period $P_\mathrm{orb}$ and a super-orbital period $P_\mathrm{sup}$.
The first BeHMXB to have shown this cycle of activity is A\,0538-66, with $P_\mathrm{orb} = 16.65$\,days and $P_\mathrm{sup} = 421$\,days \citep{mcgowan:2003}.
Then, a comprehensive study of the MACHO and OGLE light curves over 18 years light-curves of $\sim 20$ BeHMXBs show that $P_\mathrm{sup}$, extending from 500 to 4000\,days, seems to be correlated to $P_\mathrm{orb}$, from 10 to 300\,days \citep{rajoelimanana:2011}. While the origin of this variation is still unknown, this correlation shows that it is certainly due to binarity, and not intrinsic to the Be star.

\subsection{Supergiant X-ray binaries}

These systems usually host a NS on a circular orbit around an early spectral type supergiant OB donor star, with a steady wind outflow. They are separated in two distinct groups: Roche lobe overflow and wind-fed systems. The former group constitutes the classical «bright» sgHMXBs with accreted matter flowing via inner Lagrangian point to the accretion disc, causing a high X-ray luminosity ($L_X \sim 10^{38} \ergs$) during outbursts\footnote{It is interesting to note that Cyg\,X-1 is the only sgHMXB with Roche lobe overflow and stellar wind accretion hosting a confirmed BH.}.
The later group concerns close systems ($P_\mathrm{orb} < 15$\,d) with a low eccentricity, the NS accreting from deep inside the strong steady radiative and highly supersonic stellar wind. This creates a persistent X-ray emission at regular low-level effect ($L_X \sim 10^{35-36} \ergs$), exhibiting rare Type II outbursts, and no Type I outbursts. These systems exhibit large variations on short timescales, due to wind inhomogeneities.
During their long term evolution, the orbits of sgHMXBs will tend to circularize more rapidly with time, while the rate of mass transfer steadily increases \citep{kaper:2004}.

A milestone in the evolution of close binary systems takes place during what is called the ``common envelope phase''. This phase is initiated when the compact object enters inside the envelope of the companion star, in an orbit which is rapidly decreasing due to a large loss of orbital angular momentum. This phase has been invoked by \cite{paczynski:1976} to explain how high energy binary systems with very short $P_\mathrm{orb}$ can be formed, while both components of these systems -- large stars at their formation -- would not have been able to fit inside a binary system with such a small orbital separation. This phase of inward spiralling, currently taken into account in population synthesis models, but never observed yet, probably because it is short \cite[models predict a maximum duration of common envelope phase of only $\sim 1000$\,years;][]{meurs:1989} compared to the lifetime of a massive star ($\sim 10^{6-7}$\,years), is an ingredient of prime importance for understanding the evolution of high energy binary systems \citep{tauris:2006}.

\subsection{The Corbet diagram} \label{section:corbet}

\begin{figure}
\centerline{\includegraphics[width=7.cm]{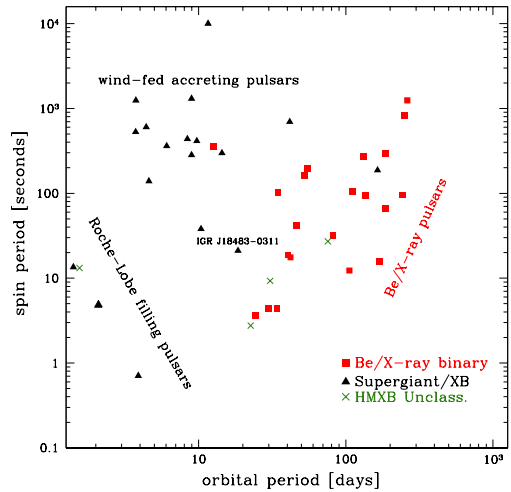}}
\caption{\label{corbet} Corbet diagram (NS $P_\mathrm{spin}$ versus system $P_\mathrm{orb}$), showing the three HMXB populations: {\it i.} systems hosting Be stars (BeHMXBs), {\it ii.} supergiant stars overflowing their Roche lobe, and {\it iii.} systems accreting the stellar wind of supergiant stars (sgHMXBs). The distinct location of these three types of HMXBs results from the interaction and equilibrium reached between accretion of matter and NS spin, altogether with the general characteristics of binary systems (such as mass ratio and orbital separation). 
The ``misplaced'' SFXT IGR\,J18483-0311 is indicated (credit J.A. Zurita Heras). 
}
\end{figure}

In case of an accreting NS in an HMXB, it is possible to detect a pulsation, corresponding to the NS spin period $P_\mathrm{spin}$. These X-ray accretion-powered pulsars are divided according to their nature and the dominant accretion process -- BeHMXBs (disc-accreting systems) or sgHMXBs (wind-accreting and Roche lobe overflow systems) -- in distinct locations in the Corbet diagram, representing the NS $P_\mathrm{spin}$ versus the system $P_\mathrm{orb}$ \cite[][see Figure \ref{corbet}]{corbet:1986}. This diagram is a valuable tool to study the interaction and feedback between the NS and accreted matter, and the influence of the local absorbing matter, the location of the different systems being determined by the equilibrium period reached by the rotation of the NS accreting matter on its surface.

It is clear from this diagram that a strong correlation exists for BeHMXBs -- $P_\mathrm{spin}\propto (P_\mathrm{orb})^2$ \citep{corbet:1984} --, due to the efficient transfer of angular momentum when the NS accretes material from the Be star decretion disc. Intuitively, a small (wide) orbit presents on average a high (low) stellar wind density, increasing (decreasing) the accretion pressure, and thus accelerating (slowing) the NS rotation, which in turn increases (decreases) the centrifugal inhibition, preventing (allowing) more accretion of matter. 
In contrast, the lack of correlation for sgHMXBs suggests that in this case, wind accretion is very inefficient to transfer angular momentum.

High energy properties, and in particular the accretion efficiency, of a compact object accreting from a circumstellar disc, a dust cocoon or the stellar wind of the companion star, will strongly differ, depending on the geometry of the system, altogether with the size and shape of the optically thin and fully ionized region of gas, similar to the Str\"omgren sphere \cite[see e.g.][]{pringle:1973, hatchett:1977}.
In particular, $P_\mathrm{spin}$ in HMXBs is determined by the stellar wind characteristics. sgHMXBs exhibit a spherically-symmetric radiation-driven wind, with a density $\rho \propto r^{-2}$ and a velocity $\sim 600-900\kms$, while BeHMXBs present an equatorial stellar wind with a density $\rho \propto r^{-3 \mbox{ to } -3.5}$ dropping faster, and a lower velocity $\sim 200-300 \kms$ \citep{waters:1988}. Therefore, larger density and velocity gradients at the distance of the NS make the wind-fed accretion transfer of angular momentum more efficient in BeHMXBs than sgHMXBs \citep{waters:1989}.

Accretion can occur on a magnetized NS only if the pressure of the infalling material is greater than the centrifugal inhibition (corresponding to the Alfven radius located inside the magnetospheric boundary). The NS should then reach an equilibrium rotation period $P_\mathrm{eq}$ corresponding to the corotation velocity $V_C$ (at the magnetospheric radius) equal to the Keplerian speed $V_K$. If $V_C > V_K$ (corresponding to $P_\mathrm{spin} < P_\mathrm{eq}$) the Propeller mechanism increases $P_\mathrm{spin}$ by ejecting material, taking away angular momentum \citep{illarionov:1975}. If $V_C < V_K$ (corresponding to $P_\mathrm{spin} > P_\mathrm{eq}$) the accretion either reduces or increases $P_\mathrm{spin}$ (spinning up or down), depending on the direction of angular momentum with respect to the NS spin. 

Taking into account the density of the stellar wind, and assuming a steady accretion rate with the angular momentum of same direction than the NS spin, $P_\mathrm{spin}$ should reach $P_\mathrm{eq} \propto \rho^{-3/7}$ \citep{waters:1988}. However, this is not the case, neither for BeHMXBs nor for sgHMXBs. $P_\mathrm{spin}$ of BeHMXBs does not correspond to $P_\mathrm{eq}$, constantly adjusting to the changing conditions in the wind, reflecting values of earlier evolution stage \citep{king:1991}. Current $P_\mathrm{spin}$ of NS in sgHMXBs is longer than predicted, closer to $P_\mathrm{eq}$ of a NS embedded in the stellar wind, while the companion star was still a main sequence star of spectral type O \citep{waters:1989}.

\section{The $\gamma$-ray sky seen by the {\it INTEGRAL} observatory}
The {\it INTEGRAL} observatory is an ESA satellite launched on 17 October
2002 by a PROTON rocket on an eccentric orbit. It hosts 4
instruments: 2 $\gamma$-ray coded-mask telescopes -- the imager IBIS
and the spectro-imager SPI, observing in the range 10 keV-10 MeV, with
a resolution of $12\amin$ and a field-of-view of $19\adeg$ --, a
coded-mask telescope JEM-X (3-100 keV), and an optical telescope
(OMC).
%
%
The $\gamma$-ray sky seen by {\it INTEGRAL} is very rich, since 723 sources
have been detected by {\it INTEGRAL}, reported in the $4^{th}$ IBIS/ISGRI soft
$\gamma$-ray catalogue, spanning nearly 7 years of observations in the 17-100
keV domain \citep{bird:2010}\footnote{See an up-to-date list at {\em http://irfu.cea.fr/Sap/IGR-Sources/}, maintained by J. Rodriguez and A. Bodaghee}.  
%
Among these sources, there are 185 X-ray binaries (representing 26\% of the whole sample of sources detected by {\it INTEGRAL}, called ``IGRs'' in the following), 255 Active Galactic Nuclei (35\%), 35 Cataclysmic Variables (5\%), and $\sim 30$ sources of other type (4\%): 15 SNRs, 4 Globular Clusters, 3 Soft $\gamma$-ray Repeaters, 2 $\gamma$-ray bursts, etc. 
215 objects still remain unidentified (30\%).
X-ray binaries are separated into 95 LMXBs and 90 HMXBs, each category representing $\sim 13$\% of IGRs. Among identified HMXBs, there are 24 BeHMXBs and 19 sgHMXBs, representing respectively 31\% and 24\% of HMXBs.

It is interesting to follow the evolution of the ratio between BeHMXBs and sgHMXBs. During the pre-{\it INTEGRAL} era, HMXBs were mostly BeHMXBs. For instance, in the catalogue of 130 HMXBs by \citet{liu:2000}, there were 54 BeHMXBs and 5 sgHMXBs (respectively 42\% and 4\% of the total number of HMXBs). Then, the situation changed drastically with the first HMXBs identified by {\it INTEGRAL}: in the catalogue of 114 HMXBs (+128 in the Magellanic Clouds) of \citet{liu:2006}, 60\% of the total number of HMXBs were firmly identified as BeHMXBs and 32\% as sgHMXBs. Therefore, while the ratio of BeHMXBs/HMXBs increased by a factor of 1.5, the sgHMXBs/HMXBs ratio increased by a larger factor of 8.
The ISGRI energy range ($> 20$\,keV), immune to the absorption that prevented the discovery of intrinsically absorbed sources by earlier soft X-ray telescopes, allowed us to go from a study of individual sgHMXBs (such as GX\,301-2, 4U\,1700-377, Vela\,X-1, etc.) to a comprehensive study of the characteristics of a whole population of HMXBs...




The most important result of {\it INTEGRAL} to date is the discovery of many new high energy sources -- concentrated in the Galactic plane, mainly towards tangential directions of Galactic arms, rich in star forming regions --, exhibiting common characteristics which previously had rarely been seen (see e.g. \citeauthor{chaty:2005a} \citeyear{chaty:2005a}). Many of them are HMXBs hosting a NS orbiting an OB companion, in most cases a supergiant star. Nearly all the {\it INTEGRAL} HMXBs for which both $P_\mathrm{spin}$ and $P_\mathrm{orb}$ have been measured are located in the upper part of the Corbet diagram (see Figure \ref{corbet}). They are X-ray wind-accreting pulsars typical of sgHMXBs, with longer pulsation periods and higher absorption (by a factor $\sim4$) compared to previously known sgHMXBs \citep{bodaghee:2007}. 
They divide into two classes: some are very obscured, exhibiting a huge intrinsic and local extinction, -- the most extreme example being the highly absorbed source IGR~J16318-4848 \citep{filliatre:2004} --, and the others are sgHMXBs exhibiting fast and transient outbursts -- an unusual characteristic among HMXBs --.  These are therefore called Supergiant Fast X-ray Transients (SFXTs, \citeauthor{negueruela:2006a} \citeyear{negueruela:2006a}),
with IGR~J17544-2619 being their archetype \citep{pellizza:2006}.

\subsection{Obscured HMXBs: IGR~J16318-4848, an extreme case}

  IGR~J16318-4848 was the first source discovered by IBIS/ISGRI on
  {\it INTEGRAL} on 29 January 2003 \citep{courvoisier:2003}, with a
  $2 \amin$ uncertainty.  {\it XMM-Newton} observations revealed a
  comptonised spectrum exhibiting an unusually high level of
  absorption: $\nh \sim 1.84 \times 10^{24} \cmmoinsdeux$
  \citep{matt:2003}.  The accurate localisation by {\it XMM-Newton}
  allowed \citet{filliatre:2004} to rapidly trigger ToO photometric and
  spectroscopic observations in optical/infrared, leading to the
  confirmation of the optical counterpart \citep{walter:2003} and to
  the discovery of the infrared one \citep{filliatre:2004}.  The extremely
  bright infrared source (B\,$>25.4\pm1$; I\,$=16.05\pm0.54$; J\,$= 10.33\pm 0.14$;
  H\,$=8.33\pm 0.10$ and K$_{\mathrm s} =7.20 \pm 0.05 \mags$) exhibits an
  unusually strong intrinsic absorption in the optical ($A_{\mathrm V} = 17.4
  \mags$), 100 times stronger than the interstellar absorption along
  the line of sight ($A_{\mathrm V} = 11.4 \mags$), but still 100 times lower
  than the absorption in X-rays.  This led \citet{filliatre:2004} to
  suggest that the material absorbing in X-rays was concentrated
  around the compact object, while the one absorbing in
  optical/infrared was enshrouding the whole system.  

Near-infrared spectroscopy in the $0.95-2.5 \microns$ domain allowed \cite{filliatre:2004} to identify the nature of the companion star, by revealing an unusual spectrum, with many strong emission lines:
%
%
H and He~I (P-Cyg) lines characteristic of dense/ionised wind at $v = 400$\,km/s,
He~II lines signatures of a highly excited region,
$[$Fe~II$]$ lines reminiscent of shock heated matter,
Fe~II lines emanating from media of densities $>10^5-10^6$\,cm$^{-3}$, and
Na~I lines coming from cold/dense regions.
%
%
All these lines originate from a highly complex, stratified
circumstellar environment of various densities and temperatures,
suggesting the presence of an envelope and strong stellar outflow/wind
responsible for the absorption. Only luminous early spectral type stars such as
sgB[e] show such extreme environments, and
\citet{filliatre:2004} concluded that IGR~J16318-4848 was an unusual
HMXB hosting a sgB[e] with characteristic luminosity $10^6 \Lsol$, mass $30 \Msol$, radius $20 \Rsol$ and temperature $T=20250$\,K, located at a distance between 1 and 6 kpc (see also \citeauthor{chaty:2005a} \citeyear{chaty:2005a}).
This source is therefore the second HMXB hosting a sgB[e] star,
after CI Cam \citep{clark:1999}. 

\begin{figure}
  \centerline{\includegraphics[width=10.6cm]{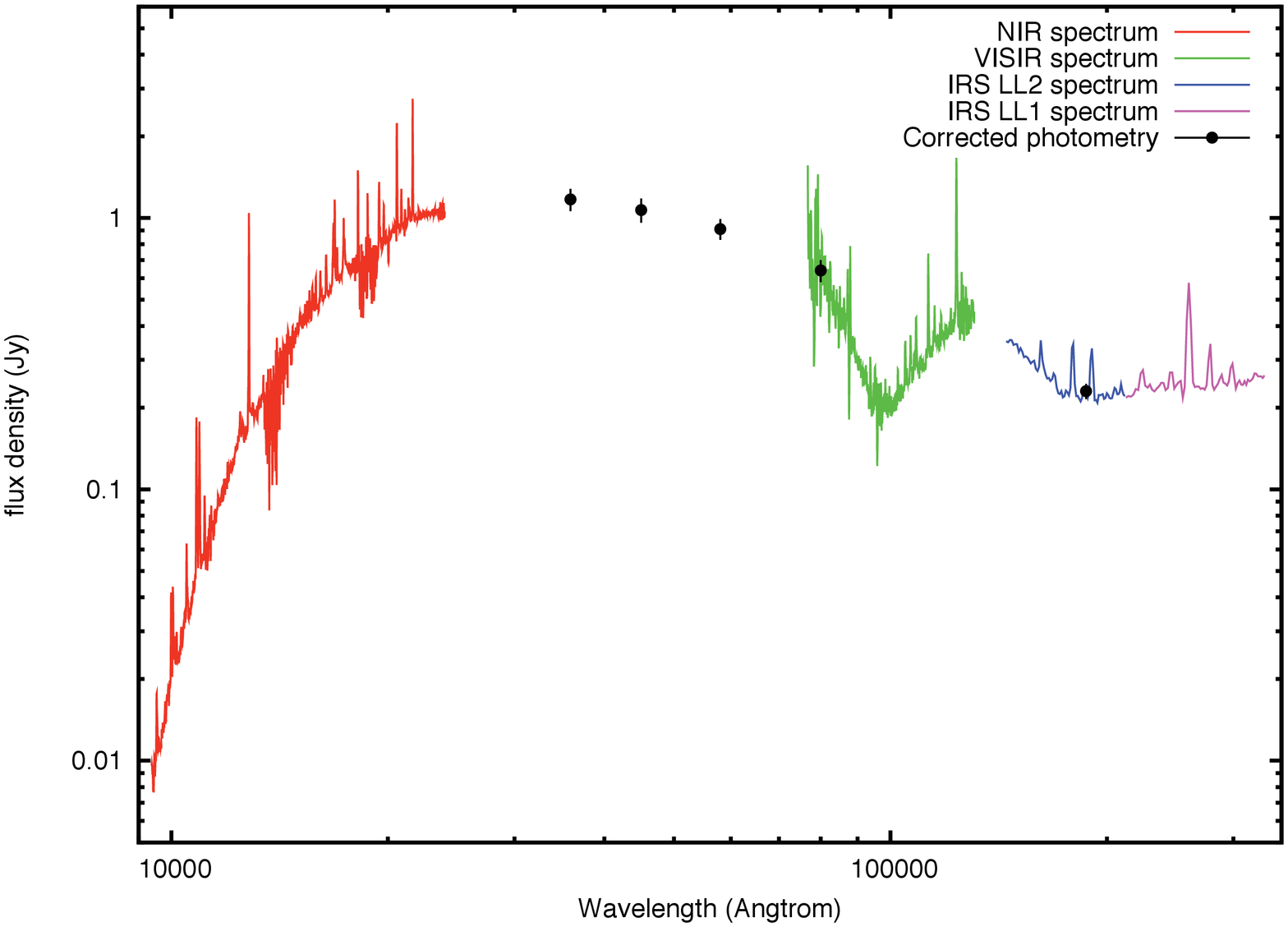}}
\caption{\label{figure:16318}
  Broadband near- to mid-infrared ESO/NTT+VISIR and {\it Spitzer} reddened spectrum of
$\igrjstu$, from 0.9 to 35 $\mu$m \citep[from][]{chaty:2011b}.
}
\end{figure}

By combining optical to mid-infrared (MIR) observations, and fitting these observations with a simple spherical blackbody, \citet{rahoui:2008} showed that IGR~J16318-4848 exhibited a MIR excess, 
interpreted as being due to the strong stellar outflow emanating from the sgB[e] companion star.  They found that this star had a temperature of $\Tstar=22000$\,K and radius $\Rstar = 20.4 \Rsol = 0.1$\,a.u., consistent with a supergiant star, with an extra component of temperature T $=1100$\,K and radius R\,$= 11.9\Rstar = 1$\,a.u., with A$_{\mathrm V} = 17.6 \mags$.
%
%
Recent MIR spectroscopic observations with VISIR at the VLT and {\it Spitzer} showed that the source was exhibiting strong emission lines of H, He, Ne, PAH, Si, proving that the extra absorbing component was made of dust and cold gas \citep{chaty:2011b}. By fitting the optical to MIR spectra with a more sophisticated aspheric disc model developped for HAEBE objets, and adapted to sgB[e] stars, they showed that the supergiant star was surrounded by a hot rim of dust at $5500$\,K, with a warm dust component at $900$\,K around it \citep[][see Figure \ref{figure:16318}]{chaty:2011b}.

By assuming a typical $P_\mathrm{orb}$ of 10\,days and a mass of the companion star of $20 \Msol$, we obtain an orbital separation of $50 \Rsol$, smaller than the extension of the extra component of dust/gas ($= 240 \Rsol$), suggesting that this dense and absorbing circumstellar material envelope enshrouds the whole binary system, like a cocoon of dust (see Figure \ref{figure:obscured-sfxt}, left panel).
We point out that this source exhibits such extreme characteristics that it might not be fully representative of the other obscured sources.

\subsection{Supergiant Fast X-ray Transients}


SFXTs constitute a new class of $\sim 12$ sources identified among the recently discovered IGRs. They are HMXBs hosting NS orbiting sgOB companion stars, exhibiting peculiar characteristics compared to ``classical'' HMXBs: rapid outbursts lasting only for hours, faint quiescent emission, and high energy spectra requiring a BH or NS accretor. The flares rise in tens of minutes, last for $\sim$ 1 hour, their frequency is $\sim7$\,days, and their luminosity $L_X$ reaches $\sim 10^{36} \ergs$ at the outburst peak.
SFXTs can be divided in two groups, according to the duration and frequency of their outbursts, and their $\frac{L_\mathrm{max}}{L_\mathrm{min}}$ ratio. ``Classical'' SFXTs exhibit a very low quiescence $L_X$ and a high variability, while ``intermediate'' SFXTs exhibit a higher average $L_X$, a lower $\frac{L_\mathrm{max}}{L_\mathrm{min}}$ and a smaller variability factor, with longer flares.  SFXTs might appear like persistent sgHMXBs with $<L_X>$ below the canonical value of $\sim 10^{36} \ergs$, and superimposed flares \citep{walter:2007}, but there might be some observational bias in these general characteristics, therefore the distinction between SFXTs and sgHMXBs is not yet well defined.
While the typical hard X-ray variability factor (the ratio between outburst flux and deep quiescence) is less than 20 in classical/absorbed systems, it is higher than 100, reaching $10^4$ for some SFXTs (some sources can exhibit flares in a few minutes, like for instance XTE\,J1739-302 \& IGR\,J17544-2619, see e.g. \citeauthor{pellizza:2006} \citeyear{pellizza:2006}).

To explain the emission of SFXTs in the context of sgHMXBs, \citet{negueruela:2008} invoked the existence of two zones in the stellar wind created by the supergiant star, of high and low clump density. This would naturally explain both the X-ray lightcurves, each outburst being due to the accretion of single clumps, and the smooth transition between sgHMXBs and SFXTs, and the existence of intermediate systems; the main difference between classical sgHMXBs and SFXTs being in this scenario the NS orbital radius.
Indeed, a basic model of porous wind with macro-clumping \citep{negueruela:2008} predicts a substantial change in the properties of the wind ``seen by the NS'' at a distance $r \sim 2 \Rstar$, where we stop seeing persistent X-ray sources. There are 2-regimes:
at $r < 2 \Rstar$ the NS sees a large number of clumps, embedded in a quasi-continuous wind; and at $r > 2 \Rstar$ the density of clumps is so low that the NS is effectively orbiting in an empty space.
NS in classical sgHMXBs can only lie within $2 \Rstar$ of the companion star.

However there exists other models to explain these accretion processes, like the formation of transient accretion discs \citep{ruffert:1996, ducci:2010} and the accretion with centrifugal/magnetic barriers \citep{bozzo:2008}.

\subsubsection{The intermediate SFXT IGR\,J18483-0311}

X-ray properties of this system were suggesting an SFXT nature
\citep{sguera:2007}, exhibiting however an unusual behaviour: its
outbursts last for a few days (to compare to hours for classical
SFXTs), and the ratio $L_{\mathrm max}/L_{\mathrm min}$ only reaches $\sim 10^3$ (compared to $\sim 10^4$ for classical SFXTs), therefore suggesting a high level quiescence. Moreover, $P_\mathrm{orb} = 18.5$\,days is low compared to classical SFXTs with large and eccentric orbits. Finally, $P_\mathrm{orb}$ and $P_\mathrm{spin} = 21.05$\,s values locate it inbetween Be and sgHMXBs in the Corbet diagram (see Figure \ref{corbet}).
\citet{rahoui:2008a} identified the companion star of this system
as a B0.5~Ia supergiant, unambiguously showing that this system is an
SFXT.  Furthermore, they suggested that this system could be the first
firmly identified intermediate SFXT, characterised by short, eccentric
orbit (with an eccentricity $e$ between 0.4 and 0.6),
and long outbursts: an intermediate SFXT nature would explain the unusual characteristics of this source among classical SFXTs.

\subsubsection{What is the origin of ``misplaced'' sgHMXBs?}

\citet{liu:2011} noted that there are two ``misplaced'' SFXTs in the Corbet diagram: IGR\,J11215-5952 \citep[a NS orbiting a B1 Ia star;][]{negueruela:2005b} and IGR\,J18483-0311 (described in the previous paragraph; see Figure \ref{corbet}). According to \citet{liu:2011}, these 2 SFXTs can not have evolved from normal main sequence OB-type stars, since {\it i.} their NS are not spinning at the equilibrium spin period $P_\mathrm{eq}$ of O\,V stars, {\it ii.} they have not been able to spin up after reaching $P_\mathrm{eq}$, and {\it iii.} their NS have not yet reached $P_\mathrm{eq}$ of sgHMXBs (see e.g. \citeauthor{waters:1989} \citeyear{waters:1989} and section \ref{section:corbet}). They must therefore be the descendants of BeHMXBs (i.e. hosting O-type emission line stars), after the NS has reached $P_\mathrm{eq}$, suggesting that some HMXBs exhibit two periods of accretion, the first one as BeHMXB and the second one as sgHMXB \citep{liu:2011}. This important result suggests that there must be many more such intermediate SFXTs.


\subsection{A scenario towards the Grand Unification of sgHMXBs}

\begin{figure}
\includegraphics[height=.235\textheight,angle=0]{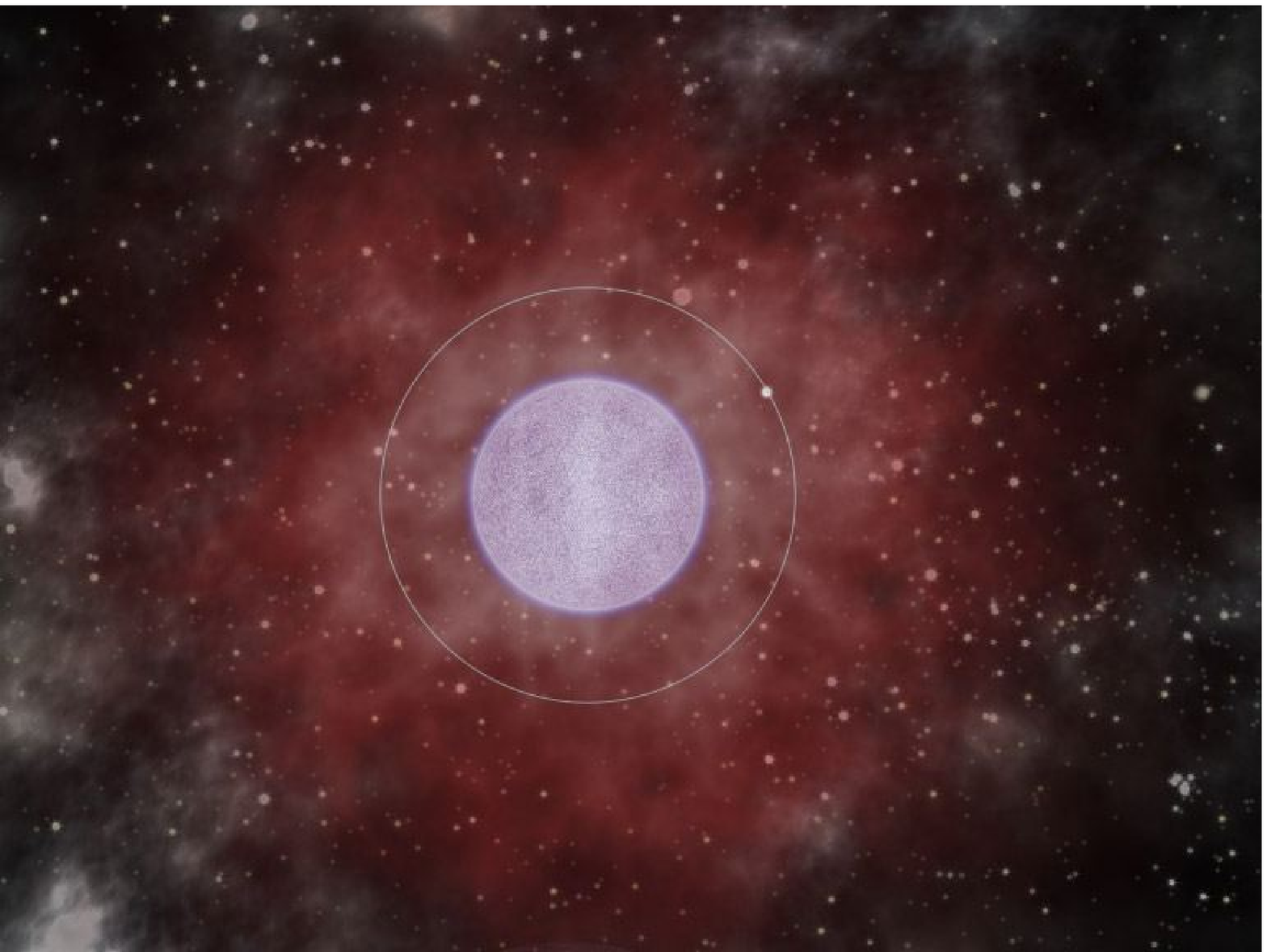}
\includegraphics[height=.235\textheight,angle=0]{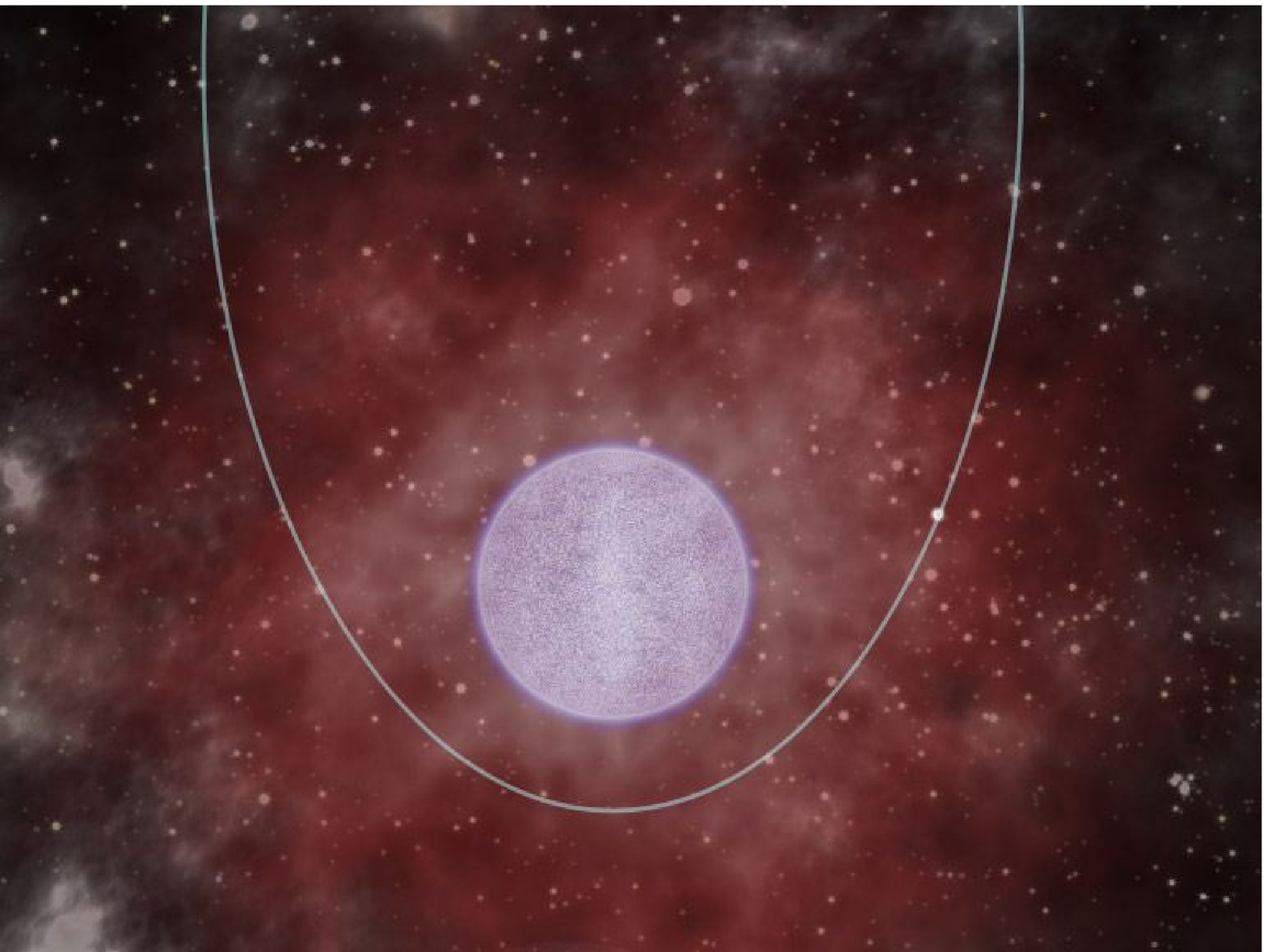}
\caption{\label{figure:obscured-sfxt}
  Scenario illustrating two possible configurations of {\it INTEGRAL} 
  sources: a NS orbiting a supergiant
  star on a circular orbit (left image); and on an eccentric orbit
  (right image), accreting from the clumpy stellar wind of the
  supergiant.  The accretion of matter is persistent in the case of
  the obscured sources, as in the left image, where the compact object
  orbits inside the cocoon of dust enshrouding the whole system. On
  the other hand, the accretion is intermittent in the case of SFXTs,
  which might correspond to a compact object on an eccentric orbit, as
  in the right image.  A 3D animation of these sources is available on the
  website: {\em http://www.aim.univ-paris7.fr/CHATY/Research/hidden.html}
}
\end{figure}

In view of the results described above, there seems to exist a continuous trend, with the observed division between classical sgHMBs and SFXTs being naturally explained by simple geometrical differences in the orbital configurations \citep[see e.g.][]{chaty:2008, chaty:2010d}:

\begin{description}

\item [Classical (or obscured/absorbed) sgHMXBs:]
  These systems (like IGR\,J16318-4848) host a NS on  a short and circular orbit at a few stellar radii only from the star, inside the zone of stellar wind high clump density ($R_\mathrm{orb} \sim 2\Rstar$). It is constantly orbiting inside a cocoon of dust and/or cold gas, probably created by the companion star itself, therefore creating a persistent and luminous X-ray emission. The cocoon, with an extension of $\sim 10 \Rstar = 1$\,a.u., is enshrouding the whole binary system (see Figure \ref{figure:obscured-sfxt}, left panel).

\item [Intermediate SFXT systems:]
In these systems (such as IGR\,J18483-0311, $P_\mathrm{orb} = 18.5$\,days), the NS orbits on a short and circular/eccentric orbit outside the zone of high clump density, and it is only when the NS penetrates inside the narrow transition zone between high and low clump density, close to the supergiant star, that accretion takes place, and that X-ray emission arises, with possible periodic outbursts.

\item [Classical SFXTs:]
In these systems (such as XTE\,J1739-302, $P_\mathrm{orb} = 50$\,days), the NS orbits outside the high density zone, on a large and eccentric orbit around the supergiant star, and exhibits some recurrent and short transient X-ray flares, when it comes close to the star, and accretes from clumps of matter coming from the wind of the supergiant.  Because it is passing through more diluted medium, the $\frac{Lmax}{Lmin}$ ratio is higher and the quiescence lasts for longer time for classical SFXTs compared to intermediate SFXTs (see Figure \ref{figure:obscured-sfxt}, right panel).

\end{description}

Although this scenario seems to describe quite well the characteristics
currently seen in sgHMXBs, we still need to identify the nature of
many more sgHMXBs to confirm it, and in particular $P_\mathrm{orb}$ and the dependance of the column density with the orbital phase of the binary system.

\section{Formation and evolution of X-ray Binaries}

\subsection{Galactic distribution of X-ray binaries}

The evidence that the stellar birthplace is very important in the formation and evolution of X-ray binaries was initially noticed examining the Galactic distribution of X-ray sources detected by {\it Ginga} \citep{koyama:1990} and {\it RXTE/ASM} \citep{grimm:2002}. Thereafter, a statistical analysis of a more complete sample of sources provided by {\it INTEGRAL} has been undertaken by \citet{lutovinov:2005a}, \citet{dean:2005} and \citet{bodaghee:2007}. These works show that LMXBs are concentrated towards the central regions of the Galaxy, and more precisely in the Galactic bulge, and then gradually decrease outside, while HMXBs, underabundant in the central kpc, are spread over the whole Galactic plane, confined in the disc, exhibiting an uneven distribution, preferentially towards the tangential directions of the spiral arms. 
This spatial distribution was expected, because LMXBs are constituted of companion stars belonging to an old stellar population, located in the Galactic bulge where they had time to migrate out of the Galactic plane ($|b| > 3-5 \adeg$). On the contrary, HMXBs -- containing young companion stars -- remain close to their stellar birthsite.

\begin{figure}
  \centerline{\includegraphics[width=11.4cm]{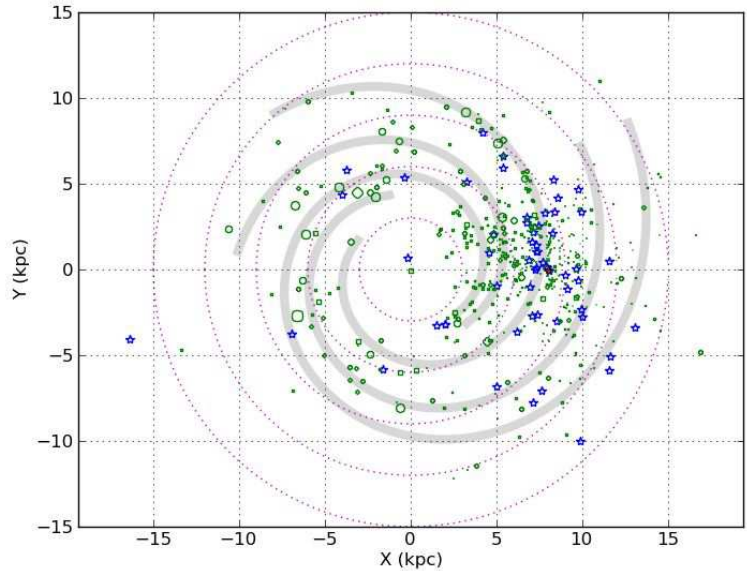}}
  \caption[Galactic distribution of HMXBs and SFCs]{Galactic distribution of HMXBs (blue stars) -- with distance derived by adjusting the spectral type on optical/infrared magnitudes of the companion star --, and of SFCs \citep[green circles, symbol size proportional to SFC activity;][]{russeil:2003}. The Sun is indicated by the red dot at (8.5, 0) (from \citeauthor{coleiro:2011} \citeyear{coleiro:2011}).
}
  \label{figure:distribution}
\end{figure}

To perform this study, one has to derive the distribution of Galactic HMXBs from the catalogues of sources detected by current high-energy satellites such as {\it XMM}, {\it Chandra}, {\it Swift} and {\it INTEGRAL}, taking into account the observational bias, some areas being more observed than others. However, the main difficulty is that most of the distances of HMXBs are not accurately determined. To overcome this difficulty, a possibility is to determine the distance of all known HMXBs, by fitting the optical and infrared magnitudes with a black body corresponding to the companion star spectral type, and comparing their distribution with catalogs of massive stars, active OB stars and star formation complexes \citep[SFCs;][]{russeil:2003}. 
A preliminary result of this method, detailed in \cite{coleiro:2011}, gives a typical clustering size of HMXBs with SFCs of 0.3\,kpc, with an average distance between these clusters of 1.7\,kpc, therefore suggesting a correlation (see Figure \ref{figure:distribution}).

The appearance of HMXBs should theoretically follow the passage of the wave density associated with the rotation of the spiral arms \citep{lin:1969}, inducing a burst of stellar formation. In this context, the models predict a time lag between star formation and HMXB apparition, depending on the position in the Galaxy, and reflecting the range of initial masses of both stars composing each binary system \citep{dean:2005}. However, since the formation of an HMXB only takes tens of million years \citep{tauris:2006}, it is a good marker for the transition from density wave in a region of the Galaxy. For such a statistical study to be accurate, we not only need many HMXBs for which we accurately know the distance, but also a precise and reliable kinematic model of the spiral arm structure of our Galaxy (Coleiro \& Chaty subm.).

\subsection{Link with population synthesis models}

sgHMXBs revealed by {\it INTEGRAL} will allow us to better understand and constrain the formation and evolution of X-ray binary systems, by comparing them to numerical study of LMXB/HMXB population synthesis models. For instance, these new systems might represent a precursor stage of what is known as the "common envelope phase" in the evolution of LMXB/HMXBs, when the orbit has shrunk so much that the NS begins to orbit inside the envelope of the supergiant star. In addition, many parameters do influence the various evolutions, from one system to another: differences in mass, size, orbital period, age, rotation, magnetic field, accretion type, stellar endpoints, etc. Moreover, stellar and circumstellar properties also influence the evolution of high-energy binary systems, made of two massive components likely born in rich star forming regions.

The formation and evolution of the supergiant {\it INTEGRAL} sources is mentioned in the very nice review about relativistic binaries by \citet{vandenHeuvel:2009}. $P_\mathrm{orb}$ in later evolutionary phases being linearly dependent of initial $P_\mathrm{orb}$, we can derive that currently wide $P_\mathrm{orb} \sim 100$\,days OB-supergiant {\it INTEGRAL} binaries require initial systems with $P_\mathrm{orb} \sim 10$\,days \citep{vandenHeuvel:2009}.
Systems having long $P_\mathrm{orb}$ are expected to survive the common envelope phase, and may then either end as close eccentric binary radio pulsar systems (double NS), or in some cases as BH-NS binaries.

No such system is known yet, however some of these systems might harbour a BH as the compact object. Of course NSs are easier to detect through X-ray pulsations, but, as Carl Sagan pointed it out, {\it ``absence of evidence is not evidence of absence''}. We should look for BHs orbiting supergiant companion stars in wind-accreting HMXBs, however this is only feasible through observational methods involving detection of extremely faint radial velocity displacement due to the high mass of the companion star, or through extremely accurate radio measurements that will be available in the future. On the other hand, massive stars lose so much matter during their evolution that they might always finish as NSs (see e.g. \citeauthor{maeder:2008} \citeyear{maeder:2008}). If this is the case, then such systems hosting BHs might not form at all. 

Finally, these sources are also useful to look for massive stellar progenitors, for instance giving birth to coalescence of compact objects, through NS/NS or NS/BH collisions. They would then become prime candidates for gravitational wave emitters, or even short-hard $\gamma$-ray bursts.

\section{Conclusions and perspectives}

Let us first recall the {\it INTEGRAL} legacy:

\begin{itemize}

\item The {\it INTEGRAL} satellite has quadrupled the total number of known Galactic sgHMXBs, constituted of a NS orbiting a supergiant star. Most of the new sources are slow and absorbed X-ray pulsars, exhibiting a large $\nh$ and long $P_\mathrm{spin}$ ($\sim1$\,ks).

\item The {\it INTEGRAL} satellite has revealed the existence in our Galaxy of two previously hidden populations of high-energy binary systems.
First, the SFXTs, exhibiting brief and intense X-ray flares -- with a peak flux of 1 Crab during 1--100s every $\sim 100$\,days --, which can be explained by accretion through clumpy winds.
Secondly, a previously hidden population of obscured and persistent sgHMXBs, composed of supergiant companion stars exhibiting a strong intrinsic absorption and long $P_\mathrm{spin}$, with the NS deeply embedded in the dense stellar wind, forming a dust cocoon enshrouding the whole binary system.

\end{itemize}

Apart from these observational facts, has {\it INTEGRAL} allowed us to better understand sgHMXBs and other populations of HMXBs? Do we better apprehend the accretion processes in HMXBs in general, and in sgHMXBs in particular, and what makes the fast transient flares so special, in the context of the clumpy wind model, and of the formation of transient accretion discs? 
In summary, has the increased population of supergiant HMXBs allowed us a better knowledge of these sources, compared to the ones that were already known before the launch of {\it INTEGRAL}? The answer to all these questions is probably {\it ``not yet''}, however we now have in hand more sources, and therefore more constraints to play with.
Studying these populations will provide a better understanding of the formation and evolution of short-living HMXBs, and study accretion processes. Furthermore, stellar population models now have to take these objects into account, to assess a realistic number of
high-energy binary systems in our Galaxy. 


\acknowledgements I would like to thank the organisers for a successfully organized and interesting workshop, involving many people working on different fields, in the nice Vi\~na del Mar!
I am grateful to Phil Charles, Christian Knigge and Thomas Tauris for fruitful discussions.
I would like to thank my close collaborators on the study of {\it INTEGRAL} sources: A. Bodaghee, A. Coleiro, P.A. Curran, Q.Z. Liu, I. Negueruela, L. Pellizza, F. Rahoui, J. Rodriguez, J.A. Tomsick, J.Z. Yan and J.A. Zurita Heras.
This work was supported by the Centre National d'Etudes Spatiales (CNES), based on observations obtained with MINE -- the Multi-wavelength {\it INTEGRAL} NEtwork --.


\begin{thebibliography}{}
\expandafter\ifx\csname natexlab\endcsname\relax\def\natexlab#1{#1}\fi
\expandafter\ifx\csname url\endcsname\relax
  \def\url#1{\texttt{#1}}\fi
\expandafter\ifx\csname urlprefix\endcsname\relax\def\urlprefix{URL }\fi
\providecommand{\eprint}[2][]{\url{#2}}

\bibitem[{{Bird} et~al.(2010){Bird}, {Bazzano}, {Bassani}, {Capitanio},
  {Fiocchi}, {Hill}, {Malizia}, {McBride}, {Scaringi}, {Sguera}, {Stephen},
  {Ubertini}, {Dean}, {Lebrun}, {Terrier}, {Renaud}, {Mattana}, {G{\"o}tz},
  {Rodriguez}, {Belanger}, {Walter}, \& {Winkler}}]{bird:2010}
{Bird}, A.~J., {Bazzano}, A., {Bassani}, L., {Capitanio}, F., {Fiocchi}, M.,
  {Hill}, A.~B., {Malizia}, A., {McBride}, V.~A., {Scaringi}, S., {Sguera}, V.,
  {Stephen}, J.~B., {Ubertini}, P., {Dean}, A.~J., {Lebrun}, F., {Terrier}, R.,
  {Renaud}, M., {Mattana}, F., {G{\"o}tz}, D., {Rodriguez}, J., {Belanger}, G.,
  {Walter}, R., \& {Winkler}, C. 2010, ApJSS, 186, 1.

\bibitem[{{Bodaghee} et~al.(2007){Bodaghee}, {Courvoisier}, {Rodriguez},
  {Beckmann}, {Produit}, {Hannikainen}, {Kuulkers}, {Willis}, \&
  {Wendt}}]{bodaghee:2007}
{Bodaghee}, A., {Courvoisier}, T.~J.-L., {Rodriguez}, J., {Beckmann}, V.,
  {Produit}, N., {Hannikainen}, D., {Kuulkers}, E., {Willis}, D.~R., \&
  {Wendt}, G. 2007, A\&A, 467, 585.

\bibitem[{{Bozzo} et~al.(2008){Bozzo}, {Falanga}, \& {Stella}}]{bozzo:2008}
{Bozzo}, E., {Falanga}, M., \& {Stella}, L. 2008, \apj, 683, 1031.

\bibitem[{{Charles} \& {Coe}(2006)}]{charles:2006}
{Charles}, P.~A., \& {Coe}, M.~J. 2006, Compact stellar X-ray sources.
  (Cambridge Astrophysics Series, Cambridge University Press), vol.~39, chap.
  {Optical, ultraviolet and infrared observations of X-ray binaries}, 215.

\bibitem[{{Chaty}(2006)}]{chaty:2006a}
{Chaty}, S. 2006, in Proceedings of Frontier Objects in Astrophysics and
  Particle Physics, Vulcano Workshop, May 22-27, 2006, edited by
  G.~{Giovannelli}, F. \&~{Mannocchi} (Italian Physical Society), vol.~93, 329.

\bibitem[{{Chaty} \& {Filliatre}(2005)}]{chaty:2005a}
{Chaty}, S., \& {Filliatre}, P. 2005, A\&ASS, 297, 235.

\bibitem[{{Chaty} \& {Rahoui}(2011)}]{chaty:2011b}
{Chaty}, S., \& {Rahoui}, F. 2011, ApJ subm.

\bibitem[{{Chaty} et~al.(2008){Chaty}, {Rahoui}, {Foellmi}, {Rodriguez},
  {Tomsick}, \& {Walter}}]{chaty:2008}
{Chaty}, S., {Rahoui}, F., {Foellmi}, C., {Rodriguez}, J., {Tomsick}, J.~A., \&
  {Walter}, R. 2008, A\&A, 484, 783.

\bibitem[{{Chaty} et~al.(2010){Chaty}, {Zurita Heras}, \&
  {Bodaghee}}]{chaty:2010d}
{Chaty}, S., {Zurita Heras}, J.~A., \& {Bodaghee}, A. 2010, in Procedings of Science, 8th INTEGRAL workshop, Dublin, Ireland, 27-30th Sept. 2010. ArXiv e-prints. \eprint{1012.2318}

\bibitem[{{Clark} et~al.(1999){Clark}, {Steele}, {Fender}, \&
  {Coe}}]{clark:1999}
{Clark}, J.~S., {Steele}, I.~A., {Fender}, R.~P., \& {Coe}, M.~J. 1999, A\&A,
  348, 888.

\bibitem[{{Coe}(2000)}]{coe:2000}
{Coe}, M.~J. 2000, in ASP Conf. Ser. 214: IAU Colloq. 175: The Be Phenomenon in
  Early-Type Stars, edited by M.~A. {Smith}, H.~F. {Henrichs}, \&
  J.~{Fabregat}, 656.

\bibitem[{{Coleiro} \& {Chaty}(2011)}]{coleiro:2011}
{Coleiro}, A., \& {Chaty}, S. 2011, ASP Conference Series; this volume.

\bibitem[{{Corbet}(1984)}]{corbet:1984}
{Corbet}, R.~H.~D. 1984, A\&A, 141, 91.

\bibitem[{{Corbet}(1986)}]{corbet:1986}
--- 1986, MNRAS, 220, 1047.

\bibitem[{{Courvoisier} et~al.(2003){Courvoisier}, {Walter}, {Rodriguez},
  {Bouchet}, \& {Lutovinov}}]{courvoisier:2003}
{Courvoisier}, T.~J.-L., {Walter}, R., {Rodriguez}, J., {Bouchet}, L., \&
  {Lutovinov}, A.~A. 2003, IAU Circ., 8063, 3.

\bibitem[{{Dean} et~al.(2005){Dean}, {Bazzano}, {Hill}, {Stephen}, {Bassani},
  {Barlow}, {Bird}, {Lebrun}, {Sguera}, {Shaw}, {Ubertini}, {Walter}, \&
  {Willis}}]{dean:2005}
{Dean}, A.~J., {Bazzano}, A., {Hill}, A.~B., {Stephen}, J.~B., {Bassani}, L.,
  {Barlow}, E.~J., {Bird}, A.~J., {Lebrun}, F., {Sguera}, V., {Shaw}, S.~E.,
  {Ubertini}, P., {Walter}, R., \& {Willis}, D.~R. 2005, A\&A, 443, 485.

\bibitem[{{Ducci} et~al.(2010){Ducci}, {Sidoli}, \& {Paizis}}]{ducci:2010}
{Ducci}, L., {Sidoli}, L., \& {Paizis}, A. 2010, \mnras, 408, 1540.

\bibitem[{{Filliatre} \& {Chaty}(2004)}]{filliatre:2004}
{Filliatre}, P., \& {Chaty}, S. 2004, ApJ, 616, 469.

\bibitem[{{Giacconi} et~al.(1962){Giacconi}, {Gursky}, {Paolini}, \&
  {Rossi}}]{giacconi:1962}
{Giacconi}, R., {Gursky}, H., {Paolini}, F.~R., \& {Rossi}, B.~B. 1962,
  Physical Review Letters, 9, 439.

\bibitem[{{Grimm} et~al.(2003){Grimm}, {Gilfanov}, \& {Sunyaev}}]{grimm:2003}
{Grimm}, H., {Gilfanov}, M., \& {Sunyaev}, R. 2003, \mnras, 339, 793.

\bibitem[{{Grimm} et~al.(2002){Grimm}, {Gilfanov}, \& {Sunyaev}}]{grimm:2002}
{Grimm}, H.-J., {Gilfanov}, M., \& {Sunyaev}, R. 2002, A\&A, 391, 923.

\bibitem[{{Hatchett} \& {McCray}(1977)}]{hatchett:1977}
{Hatchett}, S., \& {McCray}, R. 1977, ApJ, 211, 552.

\bibitem[{{Illarionov} \& {Sunyaev}(1975)}]{illarionov:1975}
{Illarionov}, A.~F., \& {Sunyaev}, R.~A. 1975, A\&A, 39, 185.

\bibitem[{{Kaper} et~al.(2004){Kaper}, {van der Meer}, \&
  {Tijani}}]{kaper:2004}
{Kaper}, L., {van der Meer}, A., \& {Tijani}, A.~H. 2004, in Revista Mexicana
  de Astronomia y Astrofisica Conference Series, edited by C.~{Allen}, \&
  C.~{Scarfe}, 128.

\bibitem[{{King}(1991)}]{king:1991}
{King}, A.~R. 1991, MNRAS, 250, 3.

\bibitem[{{Koyama} et~al.(1990){Koyama}, {Kawada}, {Kunieda}, {Tawara}, \&
  {Takeuchi}}]{koyama:1990}
{Koyama}, K., {Kawada}, M., {Kunieda}, H., {Tawara}, Y., \& {Takeuchi}, Y.
  1990, Nature, 343, 148.

\bibitem[{{Lin} et~al.(1969){Lin}, {Yuan}, \& {Shu}}]{lin:1969}
{Lin}, C.~C., {Yuan}, C., \& {Shu}, F.~H. 1969, ApJ, 155, 721.

\bibitem[{{Liu} et~al.(2011){Liu}, {Chaty}, \& {Yan}}]{liu:2011}
{Liu}, Q.~Z., {Chaty}, S., \& {Yan}, J. 2011, MNRAS in press.

\bibitem[{{Liu} et~al.(2000){Liu}, {van Paradijs}, \& {van den
  Heuvel}}]{liu:2000}
{Liu}, Q.~Z., {van Paradijs}, J., \& {van den Heuvel}, E.~P.~J. 2000, A\&ASS,
  147, 25.

\bibitem[{{Liu} et~al.(2006){Liu}, {van Paradijs}, \& {van den
  Heuvel}}]{liu:2006}
--- 2006, A\&A, 455, 1165.

\bibitem[{{Liu} et~al.(2007){Liu}, {van Paradijs}, \& {van den
  Heuvel}}]{liu:2007}
--- 2007, A\&A, 469, 807.

\bibitem[{{Lutovinov} et~al.(2005){Lutovinov}, {Revnivtsev}, {Gilfanov},
  {Shtykovskiy}, {Molkov}, \& {Sunyaev}}]{lutovinov:2005a}
{Lutovinov}, A., {Revnivtsev}, M., {Gilfanov}, M., {Shtykovskiy}, P., {Molkov},
  S., \& {Sunyaev}, R. 2005, A\&A, 444, 821.

\bibitem[{{Maeder} \& {Meynet}(2008)}]{maeder:2008}
{Maeder}, A., \& {Meynet}, G. 2008, in Revista Mexicana de Astronomia y
  Astrofisica Conference Series, vol.~33, 38.

\bibitem[{{Maragoudaki} et~al.(2001){Maragoudaki}, {Kontizas}, {Morgan},
  {Kontizas}, {Dapergolas}, \& {Livanou}}]{maragoudaki:2001}
{Maragoudaki}, F., {Kontizas}, M., {Morgan}, D.~H., {Kontizas}, E.,
  {Dapergolas}, A., \& {Livanou}, E. 2001, \aap, 379, 864.

\bibitem[{{Matt} \& {Guainazzi}(2003)}]{matt:2003}
{Matt}, G., \& {Guainazzi}, M. 2003, MNRAS, 341, L13.

\bibitem[{{McBride} et~al.(2008){McBride}, {Coe}, {Negueruela}, {Schurch}, \&
  {McGowan}}]{mcbride:2008}
{McBride}, V.~A., {Coe}, M.~J., {Negueruela}, I., {Schurch}, M.~P.~E., \&
  {McGowan}, K.~E. 2008, \mnras, 388, 1198.

\bibitem[{{McGowan} \& {Charles}(2003)}]{mcgowan:2003}
{McGowan}, K.~E., \& {Charles}, P.~A. 2003, \mnras, 339, 748.

\bibitem[{{Meurs} \& {van den Heuvel}(1989)}]{meurs:1989}
{Meurs}, E.~J.~A., \& {van den Heuvel}, E.~P.~J. 1989, A\&A, 226, 88.

\bibitem[{{Meyssonnier} \& {Azzopardi}(1993)}]{meyssonnier:1993}
{Meyssonnier}, N., \& {Azzopardi}, M. 1993, \aaps, 102, 451.

\bibitem[{{Negueruela}(2004)}]{negueruela:2004}
{Negueruela}, I. 2004, in Revista Mexicana de Astronomia y Astrofisica
  Conference Series, edited by G.~{Tovmassian}, \& E.~{Sion}, 20, 55.

\bibitem[{{Negueruela} \& {Coe}(2002)}]{negueruela:2002}
{Negueruela}, I., \& {Coe}, M.~J. 2002, \aap, 385, 517.

\bibitem[{{Negueruela} et~al.(2005){Negueruela}, {Smith}, \&
  {Chaty}}]{negueruela:2005b}
{Negueruela}, I., {Smith}, D.~M., \& {Chaty}, S. 2005, The Astronomer's
  Telegram, 470, 1.

\bibitem[{{Negueruela} et~al.(2006){Negueruela}, {Smith}, {Reig}, {Chaty}, \&
  {Torrej{\'o}n}}]{negueruela:2006a}
{Negueruela}, I., {Smith}, D.~M., {Reig}, P., {Chaty}, S., \& {Torrej{\'o}n},
  J.~M. 2006, in ESA Special Publication, edited by A.~{Wilson}, vol. 604, 165.

\bibitem[{{Negueruela} et~al.(2008){Negueruela}, {Torrejon}, {Reig}, {Ribo}, \&
  {Smith}}]{negueruela:2008}
{Negueruela}, I., {Torrejon}, J.~M., {Reig}, P., {Ribo}, M., \& {Smith}, D.~M.
  2008, in A Population Explosion: The Nature \& Evolution of X-ray Binaries in Diverse Environments. AIP Conference Proceedings, Volume 1010, 252.

\bibitem[{{Okazaki} \& {Negueruela}(2001)}]{okazaki:2001}
{Okazaki}, A.~T., \& {Negueruela}, I. 2001, \aap, 377, 161.

\bibitem[{{Paczynski}(1976)}]{paczynski:1976}
{Paczynski}, B. 1976, in Structure and Evolution of Close Binary Systems,
  edited by P.~{Eggleton}, S.~{Mitton}, \& J.~{Whelan}, vol.~73 of IAU
  Symposium, 75.

\bibitem[{{Pellizza} et~al.(2006){Pellizza}, {Chaty}, \&
  {Negueruela}}]{pellizza:2006}
{Pellizza}, L.~J., {Chaty}, S., \& {Negueruela}, I. 2006, A\&A, 455, 653.

\bibitem[{{Popov} et~al.(1998){Popov}, {Lipunov}, {Prokhorov}, \&
  {Postnov}}]{popov:1998}
{Popov}, S.~B., {Lipunov}, V.~M., {Prokhorov}, M.~E., \& {Postnov}, K.~A. 1998,
  Astronomy Reports, 42, 29.

\bibitem[{{Portegies Zwart}(1995)}]{portegies-zwart:1995}
{Portegies Zwart}, S.~F. 1995, \aap, 296, 691.

\bibitem[{{Pringle}(1973)}]{pringle:1973}
{Pringle}, J.~E. 1973, Nature, 243, 90.

\bibitem[{{Rahoui} \& {Chaty}(2008)}]{rahoui:2008a}
{Rahoui}, F., \& {Chaty}, S. 2008, A\&A, 492, 163.

\bibitem[{{Rahoui} et~al.(2008){Rahoui}, {Chaty}, {Lagage}, \&
  {Pantin}}]{rahoui:2008}
{Rahoui}, F., {Chaty}, S., {Lagage}, P.-O., \& {Pantin}, E. 2008, A\&A, 484,
  801.

\bibitem[{{Rajoelimanana} et~al.(2011){Rajoelimanana}, {Charles}, \&
  {Udalski}}]{rajoelimanana:2011}
{Rajoelimanana}, A.~F., {Charles}, P.~A., \& {Udalski}, A. 2011, \mnras, 413, 1600.

\bibitem[{{Rappaport} \& {van den Heuvel}(1982)}]{rappaport:1982}
{Rappaport}, S., \& {van den Heuvel}, E.~P.~J. 1982, in Be Stars, edited by
  {M.~Jaschek \& H.-G.~Groth}, vol.~98 of IAU Symposium, 327.

\bibitem[{{Reig}(2007)}]{reig:2007}
{Reig}, P. 2007, MNRAS, 255.

\bibitem[{{Ruffert}(1996)}]{ruffert:1996}
{Ruffert}, M. 1996, A\&A, 311, 817.

\bibitem[{{Russeil}(2003)}]{russeil:2003}
{Russeil}, D. 2003, A\&A, 397, 133.

\bibitem[{{Sguera} et~al.(2007){Sguera}, {Hill}, {Bird}, {Dean}, {Bazzano},
  {Ubertini}, {Masetti}, {Landi}, {Malizia}, {Clark}, \&
  {Molina}}]{sguera:2007}
{Sguera}, V., {Hill}, A.~B., {Bird}, A.~J., {Dean}, A.~J., {Bazzano}, A.,
  {Ubertini}, P., {Masetti}, N., {Landi}, R., {Malizia}, A., {Clark}, D.~J., \&
  {Molina}, M. 2007, A\&A, 467, 249.

\bibitem[{{Stanimirovic} et~al.(1999){Stanimirovic}, {Staveley-Smith},
  {Dickey}, {Sault}, \& {Snowden}}]{stanimirovic:1999}
{Stanimirovic}, S., {Staveley-Smith}, L., {Dickey}, J.~M., {Sault}, R.~J., \&
  {Snowden}, S.~L. 1999, \mnras, 302, 417.

\bibitem[{{Tauris} \& {van den Heuvel}(2006)}]{tauris:2006}
{Tauris}, T.~M., \& {van den Heuvel}, E.~P.~J. 2006, {Formation and evolution
  of compact stellar X-ray sources} (Compact stellar X-ray sources), 623.

\bibitem[{{van den Heuvel}(1983)}]{vandenheuvel:1983}
{van den Heuvel}, E.~P.~J. 1983, in Accretion-Driven Stellar X-ray Sources,
  edited by {W.~H.~G.~Lewin \& E.~P.~J.~van den Heuvel}, 303.

\bibitem[{{van den Heuvel}(2009)}]{vandenHeuvel:2009}
--- 2009, in Astrophysics and Space Science Library, edited by {M.~Colpi,
  P.~Casella, V.~Gorini, U.~Moschella, \& A.~Possenti }, vol. 359 of
  Astrophysics and Space Science Library, 125.

\bibitem[{{Verbunt} \& {van den Heuvel}(1995)}]{verbunt:1995}
{Verbunt}, F., \& {van den Heuvel}, E.~P.~J. 1995, in X-ray Binaries, edited by
  {W.~H.~G.~Lewin, J.~van Paradijs, \& E.~P.~J.~van den Heuvel}, 457.

\bibitem[{{Walter} et~al.(2003){Walter}, {Rodriguez}, {Foschini}, {de Plaa},
  {Corbel}, {Courvoisier}, {den Hartog}, {Lebrun}, {Parmar}, {Tomsick}, \&
  {Ubertini}}]{walter:2003}
{Walter}, R., {Rodriguez}, J., {Foschini}, L., {de Plaa}, J., {Corbel}, S.,
  {Courvoisier}, T.~J.-L., {den Hartog}, P.~R., {Lebrun}, F., {Parmar}, A.~N.,
  {Tomsick}, J.~A., \& {Ubertini}, P. 2003, A\&A, 411, L427.

\bibitem[{{Walter} \& {Zurita Heras}(2007)}]{walter:2007}
{Walter}, R., \& {Zurita Heras}, J. 2007, A\&A, 476, 335.

\bibitem[{{Waters} et~al.(1988){Waters}, {van den Heuvel}, {Taylor}, {Habets},
  \& {Persi}}]{waters:1988}
{Waters}, L.~B.~F.~M., {van den Heuvel}, E.~P.~J., {Taylor}, A.~R., {Habets},
  G.~M.~H.~J., \& {Persi}, P. 1988, A\&A, 198, 200.

\bibitem[{{Waters} \& {van Kerkwijk}(1989)}]{waters:1989}
{Waters}, L.~B.~F.~M., \& {van Kerkwijk}, M.~H. 1989, A\&A, 223, 196.

\end{thebibliography}

\end{document}